\documentclass[11pt,twoside]{article} 
\usepackage{cspm-asp2006} 
\usepackage{graphicx,natbib}  
\usepackage{lscape} 
\begin{document} 
\setcounter{page}{359} 
\markboth{Schwartz et al.}{Non-LTE Analysis of Filament Lyman Lines}
\title{Non-LTE Analysis of Lyman-Line Observations of a Filament 
       with SUMER}
\author{P. Schwartz$^1$, 
        B. Schmieder$^2$ 
        and 
        P. Heinzel$^1$}  
\affil{$^1$Astronomical Institute AS, Ond\v{r}ejov, Czech Republic\\
       $^2$Observatoire de Paris, Section Meudon, LESIA, Meudon, France} 

\begin{abstract}  
We present non-LTE diagnostics of the filament observed by
SOHO/SUMER on May 27, 2005 in the whole Lyman series. The filament was
situated close to the disk center. The L$\alpha$ observations were
carried out with normal voltage of detector A.  The slit was placed at
the central part of the detector -- outside the L$\alpha$
attenuator.  Therefore, the observed profiles of this line could be
calibrated reliably.
\end{abstract} 

\section{Introduction}\label{schwsec:intro}
In previous works \citep{sch-schwartz2006a, sch-schwartz2006b} we
made the non\discretionary{-}{-}{-}LTE modeling of the profiles of
L$\beta$ and higher hydrogen Lyman lines observed in EUV filaments by
SoHO/SUMER on October 15, 1999 and May 5, 2000. For the latter EUV
filament we had also observations of the H$\alpha$ profiles from
THEMIS/MSDP.  In \citet{sch-schwartz2006a} it was found that estimates
the of temperature in the filament interior need not be reliable if
there are temperatures lower than 10\,000\,K in the filament interior
and rather hot \hbox{PCTRs} (prominence-corona transition regions)
with temperatures above 20\,000\,K at the same time. These problems
occur especially for the H$\alpha$ filaments.  This problem can be
solved by constraining the models with the profile of the H$\alpha$
line that is not sensitive to the high temperature plasma of
\hbox{PCTRs} \citep{sch-schwartz2006b}.  Therefore its shape represent
well the temperature structure of the cool filament interior. However,
the problem is to find suitable H$\alpha$ observations with the same
position of the slit and made at the same time, as the SoHO/SUMER
observations. The optically thick cores of the Lyman lines are formed
in the top PCTRs while the optically thick parts of the wings are
formed deeper. Wavelength intervals where the profiles of L$\beta$ and
higher Lyman lines are optically thick, are small ($\sim\pm$0.2\,\AA\
or even smaller). Outside these interval the filament is
transparent. Therefore the wings of these lines do not map the
temperature structure of filament much deeper then the top PCTR. The
profile of the L$\alpha$ line could be much more sensitive to the
cooler hydrogen plasma deeper in the filament than higher Lyman
lines. Its is because of the wide wavelength interval
($\sim\pm$0.3\,--\,1.4\,\AA\ or even larger) of the optically thick
part of the profile, possibly spreading far into the wings.

In this work we are modeling profiles of the Lyman lines, Lyman
continuum and H$\alpha$ line observed in the H$\alpha$ filament using
the 1D-slab non\discretionary{-}{-}{-}LTE model \citep{sch-1dsmodcit}.
As the results of a such diagnostics we obtain the temperatures, the
gas pressure, plasma densities, ionization degree etc.

\section{Observations}\label{schwsec:observations}
A filament close to the solar disk ($\mu$=0.9) was observed on May 27,
2005 (during the 15th MEDOC observing campaign), in EUV spectral lines
by the CDS (Coronal Diagnostic Spectrometer)
\citep{sch-cdscit} and SUMER (Solar Ultraviolet Measurements
of Emitted radiation) \citep{sch-sumercit} both
on\discretionary{-}{-}{-}board of SoHO (Solar and Heliospheric
Observatory) and in the H$\alpha$ line by HSFA2 multicamera
spectrograph at Ond\v{r}ejov observatory. The SUMER observations of
the filaments and prominences during this observing campaign are
unique because the L$\alpha$ line was placed on the bare part of the
detector A (outside the attenuator) for the first time during any
filament/prominence observations and thus it was possible to make a
reliable calibration of the observed L$\alpha$ profiles.

Observations with CDS and SUMER were carried out between 17:14 and
18:07\,UT. CDS observed the EUV filament in three coronal
EUV\linebreak lines Mg\,\uppercase{x}\,624.94\,\AA,
Ca\,\uppercase{x}\,557.77\,\AA\ and Mg\,\uppercase{ix}\,368.07\,\AA,
two transition\discretionary{-}{-}{-}region EUV lines
O\,\uppercase{v}\,629.73\,\AA\ and Ne\,\uppercase{vi}\,562.80\,\AA\
and one chromospheric line \linebreak He\,\uppercase{i}\,584.33\,\AA.

\begin{figure}
\parbox{0.4\hsize}{
\resizebox{\hsize}{!}{
\includegraphics{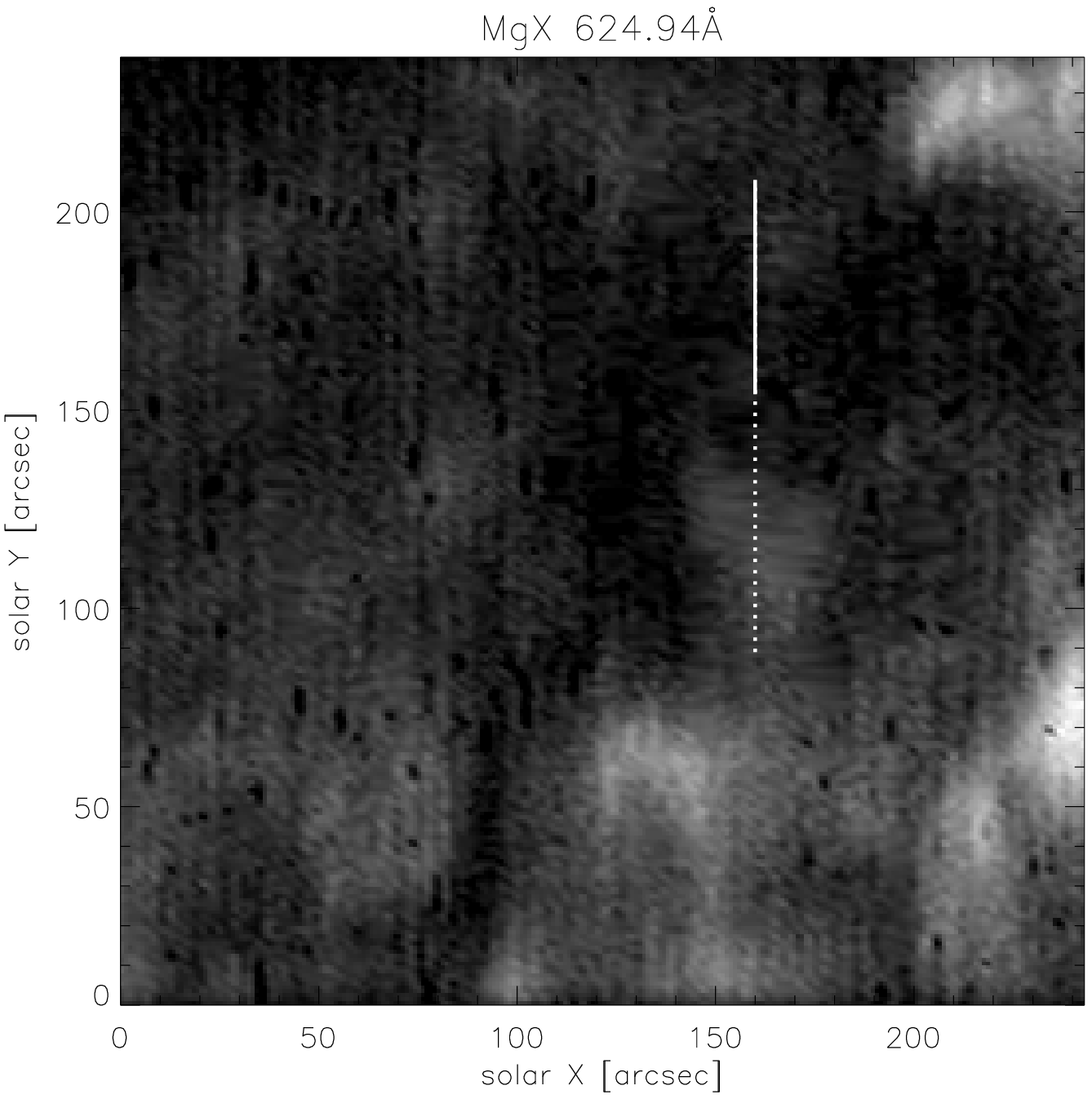}}}\
\parbox{0.4\hsize}{
\resizebox{\hsize}{!}{
\includegraphics{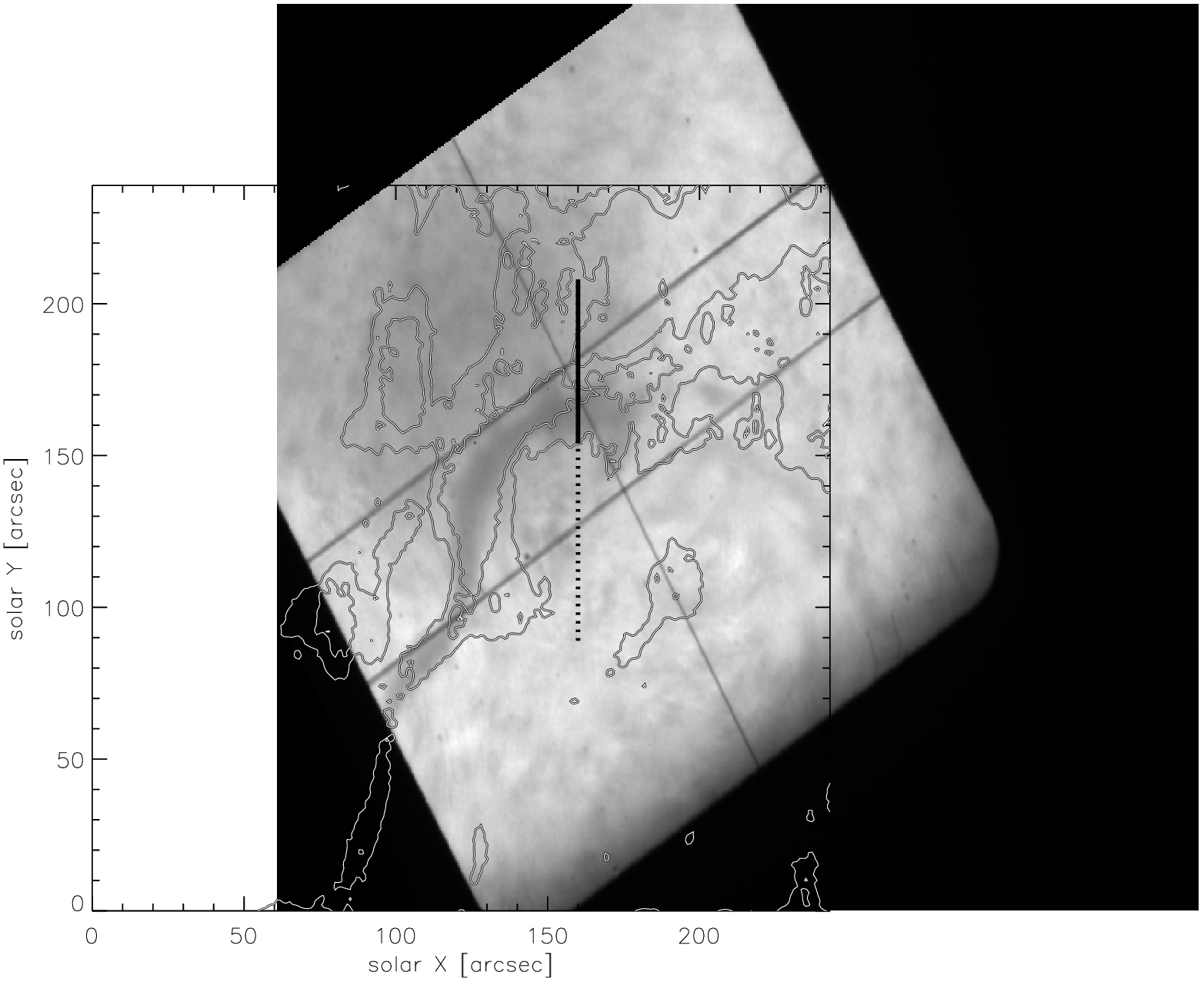}}}
\caption[]{CDS observations of the EUV filament in 
the Mg\,\uppercase{x}\,624.94\,\AA\ line are shown in the left panel. 
In the right panel the slit\discretionary{-}{-}{-}jaw of  
H$\alpha$ observations of HSFA2 spectrograph are shown. The 
slit\discretionary{-}{-}{-}jaw is co\discretionary{-}{-}{-}aligned with 
the CDS observations of the He\,\uppercase{i}\,584\,\AA\ line represented by 
contours. The inclined hair crossed perpendicularly with two other 
hairs shows the position of the spectrograph slit during the observations 
of the H$\alpha$ line. Vertical bar in both panels shows the position 
of the SUMER slit during the observations of the Lyman line series. 
Full\discretionary{-}{-}{-}line part of the bar marks working part 
of the SUMER detector A.}
\label{schfig:mgxcdsobs}
\end{figure}

Position of the center of CDS rasters is 248\,\arcsec, $-$67\,\arcsec\  
(S\,$5$ W\,$15$ in Carrington coordinates) and
their dimensions are 244\,\arcsec$\times$240\,\arcsec. CDS observations 
in the Mg\,\uppercase{x}\,624.94\,\AA\ line are shown in the left panel of 
Fig.~\ref{schfig:mgxcdsobs}. 

The H$\alpha$ observations were carried out with the HSFA2 multicamera
spectrograph of the Ond\v{r}ejov observatory at 7:14\,UT.  The
slit\discretionary{-}{-}{-}jaw co\discretionary{-}{-}{-}aligned with
the CDS observations is shown in right panel of
Fig.~\ref{schfig:mgxcdsobs}. The slit positions of HSFA2 and SUMER
spectrographs are crossing the filament in different directions and
the times of observations of HSFA2 and SoHO differs. However, since
the filament seemed to be rather stable and compact, the HSFA2
observations of H$\alpha$ could be used as additional data in our
non\discretionary{-}{-}{-}LTE modeling.

\begin{figure}
\centering
\resizebox{\hsize}{!}{
\includegraphics{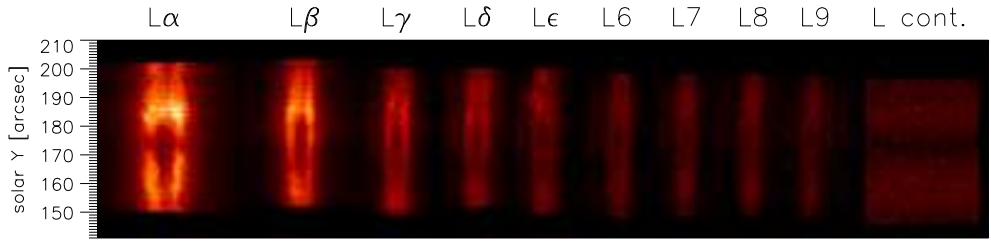}}
\caption[]{Spectra of the hydrogen Lyman lines L$\alpha$ -- 
L9 plus Lyman continuum observed by SUMER with detector A. 
The slit was positioned across 
the EUV filament as it is shown in the left panel of 
Fig.~\ref{schfig:mgxcdsobs}. Only $59\,\arcsec$ of the spectra were obtained 
because only a part of the detector was working.}
\label{schfig:sumerobs}
\end{figure}

SUMER observed the EUV filament in a wide wavelength range so that the
whole hydrogen Lyman line series is present.  The spectra of Lyman
lines L$\alpha$ -- L9 plus Lyman continuum are shown in
Fig.~\ref{schfig:sumerobs}. We do not use the lines L10 and L11
because with a 12\discretionary{-}{-}{-}level model hydrogen atom our
calculated populations of levels 11 and 12 are rather
unprecise. Position of the SUMER slit during observation of the Lyman
series is shown in Fig.~\ref{schfig:sumerobs}. Only a part of the
detector A was working therefore we obtained spectra from this part of
the slit only (marked by full\discretionary{-}{-}{-}line part of
vertical bar in both panels of Fig.~\ref{schfig:mgxcdsobs}). This
part of the slit is crossing almost only the darkest part of the EUV
filament -- the H$\alpha$ filament.

\section{Non-LTE Model of the Filament}\label{schwsec:nltemodel}
A filament is approximated by the 1D horizontal isobaric slab
\citep{sch-1dsmodcit} with temperature symmetrically decreasing from
PCTRs to the interior. The radiative transfer is solved using MALI
method \citep{sch-malicit1,sch-malicit2} and a
12\discretionary{-}{-}{-}level model of the hydrogen atom. We used
$\chi^2$ minimization method proposed in \citet{sch-schwartz2006a} for
fitting the observed profiles with synthetic ones. For reconstruction
of Lyman line profiles emitted from beneath the filament (background
irradiation) we used the method developed in \citet{sch-schwartz2006a}
-- the profiles of the background irradiation are identical with
profiles from the filament in the optically thin wings.  The optically
thick cores of L$\beta$ -- L9 lines were reconstructed using the
quiet\discretionary{-}{-}{-}Sun profiles published by
\citet{sch-warrencit}. For L$\alpha$ line the average profiles from
the quiet\discretionary{-}{-}{-}Sun observations carried out on April
14, 2005 between 13:26 and 14:26\,UT were used \citep{sch-lalpbgcit}.
During these observations the raster scan in the
quiet\discretionary{-}{-}{-}Sun area was made and the L$\alpha$ line
was placed at various positions on the bare part of the detector
A. There was a problem with the reconstruction of the
background\discretionary{-}{-}{-}irradiation profiles of the L$\alpha$
line because the width of the wavelength interval of optically thick
central part of its profile is much more sensitive to optical
thickness in the line center than it is for higher Lyman
lines. Therefore only those optically thick parts of profiles of this
line were modeled which transmit no radiation from below the filament.

\section{Results and Conclusions}\label{schwsec:resandconcl}
We obtained similar plasma properties as for other two H$\alpha$
filaments studied in \citet{sch-brigitte2003} and
\citet{sch-schwartz2006a} -- temperatures around 6\,000\,K and
20\,000\,K in the filament interior and PCTRs, respectively. PCTRs
both occupy less than 30\,\% of the geometrical thickness of the
filament, plasma densities are $10^{-14}$--
$10^{-13}\,\mathrm{g}\,\mathrm{cm}^{-3}$, electron densities around
$10^{10}\,\mathrm{cm}^{-3}$ and the hydrogen ionization degree is
lower than 0.5 in the filament interior. Only the estimated gas
pressure $\sim\!0.4\,\mathrm{dyn}\,\mathrm{cm}^{-2}$ is about 3 times
lower than that estimated for two other filaments.

We compared results of our modeling when fitting profiles of the
whole Lyman series plus Lyman continuum without H$\alpha$ and results
obtained with the H$\alpha$ line but without L$\alpha$ and found
similar plasma properties in both cases. But when fitting only Lyman
lines without L$\alpha$ the temperatures in the filament interior were
underestimated. From our analysis of the dependence of the
contribution function (computed using Eq.~(13) of
\cite{sch-contrfcit}) on the geometrical depth we found that the core
of the Lyman lines is formed at the top of PCTR, in contrast with the
H$\alpha$ line profile that is formed almost completely in the cool
interior. However, due to large optical thickness of L$\alpha$, its
near wings are formed in cool parts of the H$\alpha$ filament and this
helps to determine the temperature of the filament interior. Using the
H$\alpha$ line gives a similar result. \\

\acknowledgements This work was partly supported by grants A3003203
and 1QS300120506 of the Grant Agency of the Academy of Sciences of the
Czech Republic, institutional project AV0Z10030501, by ESA-PECS
project No.~98030 and by the European Solar Magnetism Network
(ESMN\discretionary{-}{-}{-}HPRN\discretionary{-}{-}{-}CT\discretionary{-}{-}{-}2002\discretionary{-}{-}{-}00313).
SoHO is a space mission of international cooperation between ESA and
NASA.  The SUMER data have been reduced with the intensity calibration
procedure developed at MPI Lindau and the wavelength calibration
procedure of M. Carlsson. All observations were obtained during the
15th MEDOC observing campaign.

\end{document}